\documentclass[12pt]{article}

\usepackage{geometry} 
\usepackage[numbers,sort&compress]{natbib} 

\usepackage{authblk}

\usepackage{amsmath}

\usepackage{graphicx}
\usepackage{subcaption}
\usepackage{caption}

\usepackage{url} 
\usepackage{color}
\usepackage{soul}

\title{Predicting the properties of black-hole merger remnants with Deep Neural Networks}

\author[1, 2]{L. Haegel}
\author[1]{S. Husa}
\affil[1]{Universitat de les Illes Balears, IAC3—IEEC, E-07122 Palma de Mallorca, Spain}
\affil[2]{ Université de Paris, CNRS, Astroparticule et Cosmologie, F-75013 Paris, France}

\date{} 

\begin{document}

\maketitle

\abstract{
We present the first estimation of the mass and spin magnitude of Kerr black holes resulting from the coalescence of binary black holes using a deep neural network. 
The network is trained on a dataset containing 80\% of the full publicly available catalog of numerical simulations of gravitational waves emission by binary black hole systems, including full precession effects for spinning binaries.
The network predicts the remnant black holes mass and spin with an error less than 0.04\% and 0.3\% respectively for 90\% of the values in the non-precessing test dataset, it is 0.1\% and 0.3\% respectively in the precessing test dataset.
When compared to existing fits in the LIGO algorithm software library, the network enables to reduce the remnant mass root mean square error to one half in the non-precessing case. 
In the precessing case, both remnant mass and spin mean square errors are decreased to one half, and the network corrects the bias observed in available fits.
}

\section{Introduction}
\label{intro}

General relativity predicts that binary systems of black holes (BHs) coalesce by emitting gravitational waves (GWs).
During the first two observational runs of advanced LIGO \cite{1992Sci...256..325A,TheLIGOScientific:2016agk} and advanced Virgo
 \cite{TheVirgo:2014hva}, GWs from ten binary black holes (BBHs) mergers have been detected~\cite{LIGOScientific:2018mvr}.
Current GW instruments detect solar-masses BBHs in their late inspiral and merger phase of the coalescence, resulting in a Kerr BH characterized by its final mass $M_f$ and final spin  $S_f$.
While the GW emission during the inspiral phase where the black holes separation is large can be computed in the post-Newtonian (PN) formalism \cite{Blanchet_2014}, such perturbative methods are inadequate for the merging part that must be determined using numerical relativity (NR).
Due to the high computational cost of NR simulations, a limited catalog of BBH configurations is currently available.
The information of the NR simulations must then be interpolated to cover the parameter space of the BH remnants detected by advanced LIGO and advanced Virgo.
The relation between the initial and remnant BHs properties can be determined in fits based on available NR simulations and interpolated to a wider range of parameters~\cite{Jimenez-Forteza:2016oae, Husa:2015iqa, PhysRevD.90.104004, hofmann2016final}, that are notably necessary for the development of  phenomenological fits of full gravitational waveforms~\cite{Hannam:2013oca, Khan:2015jqa}.
An accurate estimation of the remnant properties is of interest for fundamental physics, such as inspiral-merger-ringdown consistency tests aiming at testing the nature of the resulting BH~\cite{LIGOScientific:2019fpa}.
They can also provide an estimate on the remnant parameters when only the inspiral part of the waveform is detected and the merger is outside the detection range, such as the case of small-masses BBHs for LIGO and Virgo.

The existing fits rely on explicit ans\"atz that convey a good accuracy in predicting the  properties of remnants BHs in the non-precessing case, but are difficult to extend to the high number of dimensions required to describe precessing BBHs.
Consequently, the existing fits only approximate precession by parameterizing a projection of the spins onto the angular momentum instead of using the full information of the spin vectors~\cite{PhysRevD.90.104004,hofmann2016final}. 
In this article, we study the feasibility of using deep neural networks (DNN) to infer the relationship between the initial BBHs parameters and the remnant final mass $M_f$ and final dimensionless spin $\chi_f = S_f / M_f^2$ with a complete characterization of precession.
DNN are networks of nodes organized in layers, each node being assigned a weight that is propagated to the next layer according to a non-linear activation function.
The first and last layers have the dimension of the input and output parameters sizes respectively, and the weights of the DNN are adjusted with a backpropagation algorithm based on a stochastic gradient descent until the prediction is optimized.
DNN have been shown to be powerful in extracting features from large datasets with high dimensionality and complex interdependencies, enabling to deliver accurate predictions on new data~\cite{lecun2015deep}.

Motivated by those features, we train a DNN to estimate the remnant properties in precessing BBHs systems with the goal of comparing the predictions with the fits available in the LIGO software library used by the LIGO-Virgo collaboration.
They include fits for precessing configurations based on aligned-spin configurations augmented by the in-plane spin component, as well as fits in non-precessing configurations for which we train a second DNN for comparison. 
Section~\ref{inputs} describes the NR and extreme mass ratio data used as input of the DNN, while the fit procedure and results are given on Section~\ref{fits} and discussed on Section~\ref{conclusion}.
In addition to the interest in accurate models for the final state of BBH mergers, the present work can serve as an example of how to model other quantities of interest, and in principle the whole waveform, directly or in terms of the coefficients of some phenomenological model.
Our work is also motivated by the fact that Gaussian processes have shown to be powerful tools in estimating the remnant BH properties from the full description of the initial parameters~\cite{Varma:2018aht}.

\section{Input data}
\label{inputs}

\subsection{Numerical relativity catalogs}
\label{nr}

The DNN are trained and tested with data from catalogs of NR simulations of the GW emitted by BBHs, in which the initial BHs are characterized by their individual masses $m_1$ and $m_2$ with the convention $m_1 > m_2$. 
Since the total BBH mass acts as a simple scale factor in GR, we choose $M = m_1 + m_2 = 1$ for simplicity, and we parameterize the masses by the symmetric mass ratio $\eta = (m_1 m_2) / (m_1 + m_2)^2$.
For the individual spins we introduce the dimensionless spins $\chi_{1,2} =  S_{1,2} / m_{1,2}^2$ where $S_1$ and $S_2$ are the individual angular momenta. 
The remnant BH is defined by its final mass $M_f$ and spin $\chi_f$.

\begin{figure}[ht!]
      \captionsetup{width=.85\textwidth}
      \captionsetup{font=small}
	\centering
    \begin{subfigure}[t]{0.4\textwidth}
        \centering
        \includegraphics[width=\textwidth]{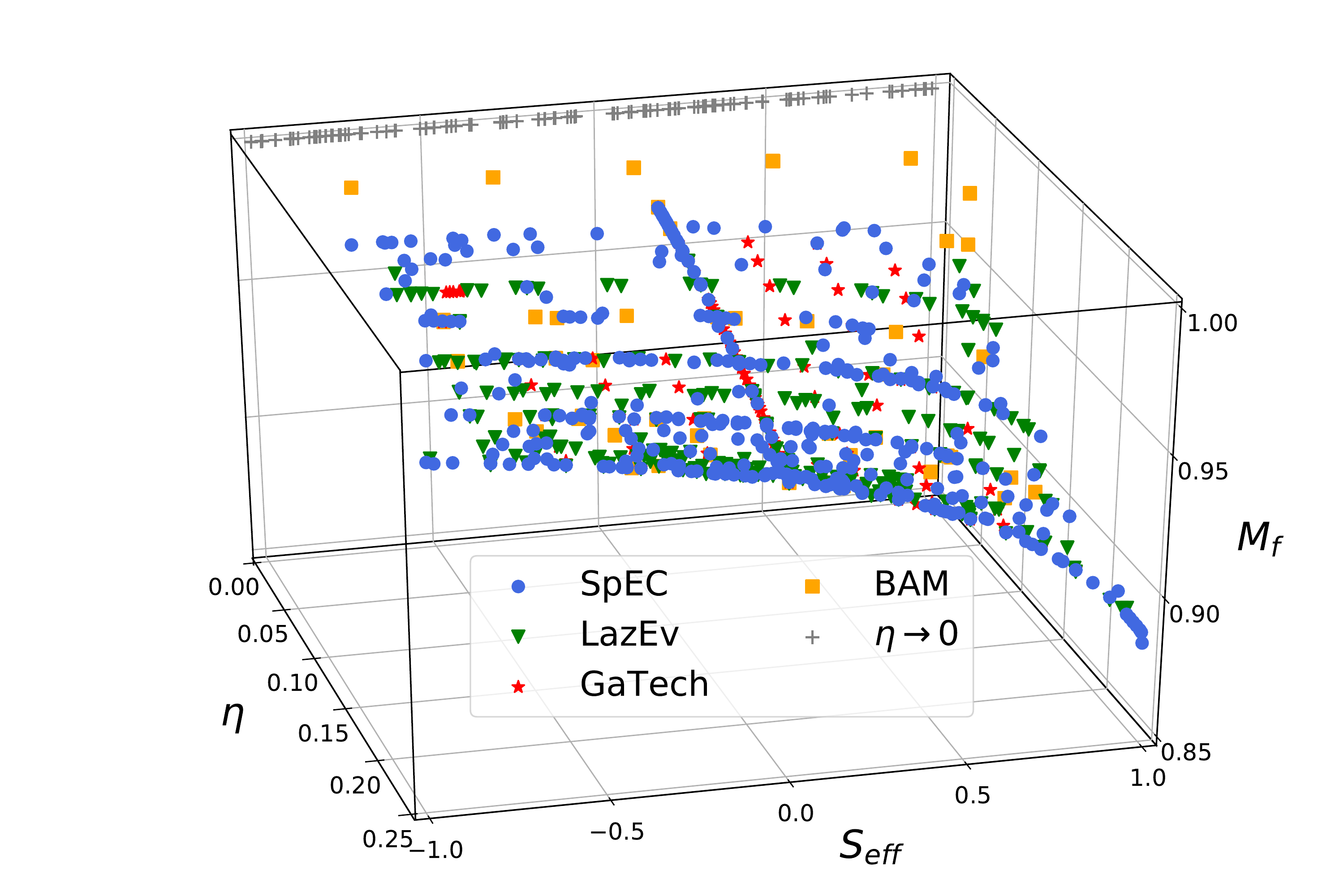}
        \caption{$M_f(\eta, S_{eff})$}
    \end{subfigure}%
    ~ 
    \begin{subfigure}[t]{0.4\textwidth}
        \centering
        \includegraphics[width=\textwidth]{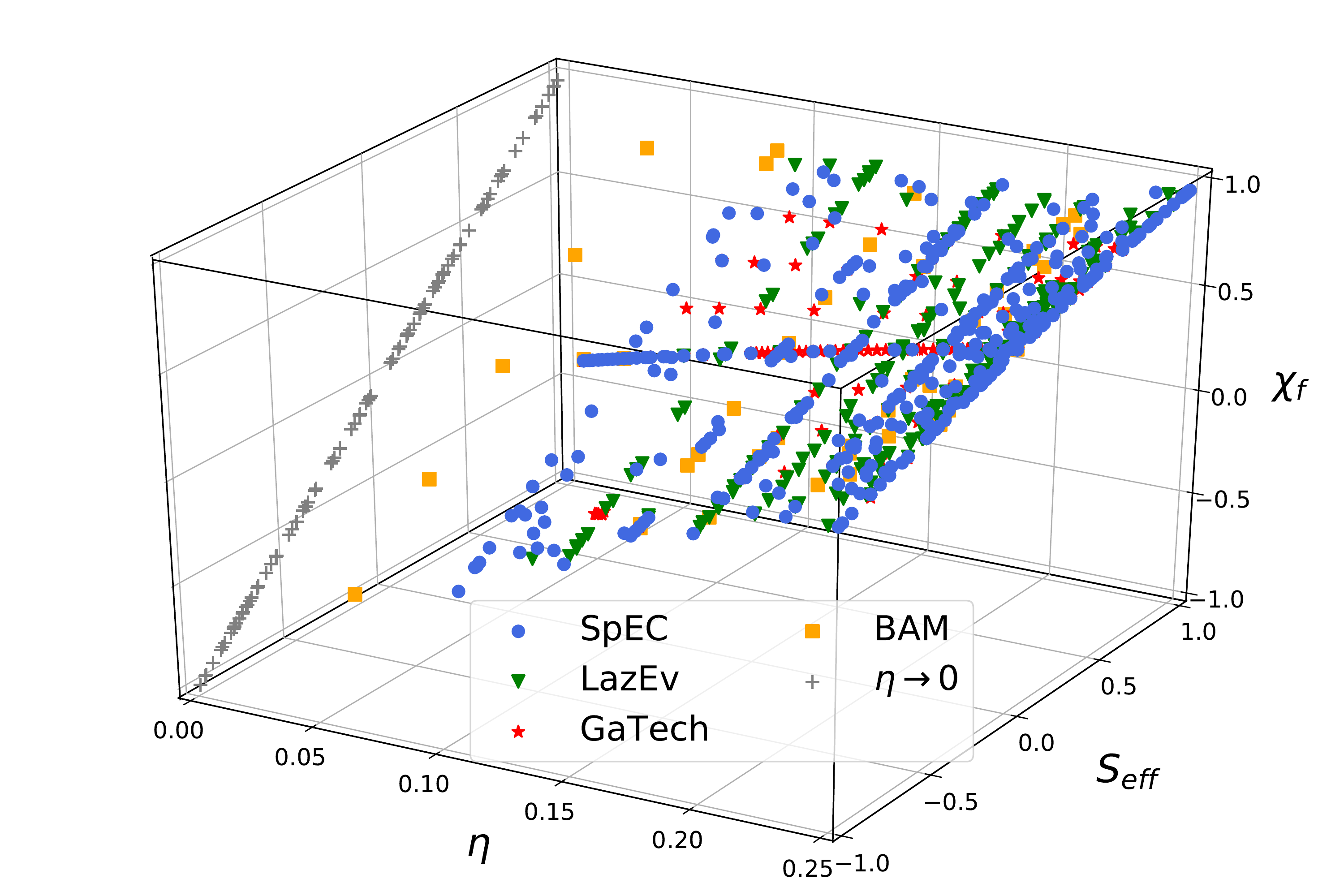}
        \caption{$\chi_f(\eta, S_{eff})$}
    \end{subfigure}
    \caption{Final properties of the remnant BH as a function of $\eta$ and $S_{eff}$ for the non-precessing cases of Table~\ref{tab:nrdata}. The $\eta \rightarrow 0$ case is described in Section~\ref{emr}. }
   \label{fig:nrdata}
\end{figure}

DNNs are known to perform better when trained with large datasets, therefore all the publicly available NR catalogs have been used, i.e. from the SpEC~\cite{SXS:catalog,  Chu:2009md, Lovelace:2010ne, Lovelace:2011nu, Buchman:2012dw, Mroue:2012kv, Hemberger:2013hsa, Hinder:2013oqa,  Mroue:2013xna, Lovelace:2014twa, Scheel:2014ina, Blackman:2015pia,  Kumar:2015tha, Chu:2015kft, Lovelace:2016uwp, Abbott:2016nmj, Abbott:2016apu, Abbott:2016wiq, Varma:2018mmi, Varma:2018aht, Boyle_2019}, LaZev~\cite{PhysRevD.90.104004, PhysRevLett.114.141101, PhysRevD.91.104022, PhysRevD.93.044031, PhysRevD.93.124074, PhysRevD.95.024037, PhysRevD.96.024031, Healy_2017, PhysRevD.97.064027, PhysRevD.97.104026, PhysRevD.97.084002}, MayaKranc (GaTech)~\cite{Jani_2016} and BAM~\cite{Bruegmann:2006at, Husa:2007hp} codes. 
We select the non-precessing sample as including no-spin and aligned-spins binaries, totaling in 1044 simulations as summarized in Table~\ref{tab:nrdata}.
We follow \cite{Husa:2015iqa} in parameterizing the 2-dimensional spin parameter space by a dominant (``effective'') spin parameter $S_{eff} = (S_1 + S_2)/(1-2 \eta M^2)$ with the property $-1 \leq S_{eff} \leq 1$, and the spin difference  $\Delta \chi = \chi_1 - \chi_2$.
The following parameter ranges are covered: $\eta \in [0.050, 0.250]$, $S_{eff} \in [-0.970, 0.999]$, $\Delta \chi \in [-1.900, 1.861]$, $M_f \in [0.883, 0.997]$, $\chi_f \in [-0.527, 0.951]$.
The distributions of the remnant parameters as a function of the symmetric mass ratio and effective spin is shown in Figure~\ref{fig:nrdata}.

The precessing sample is selected from the relevent configurations in the SpEC catalog and consists in 1420 NR simulations.
In order to fully characterize the precession, we use as input data for the DNN the asymmetric mass ratio $\eta$ as well as the two spin vectors $\overrightarrow{\chi_{1,2}}$.
The spins are determined closed to the innermost stable orbit at $t=-100M$ in the co-orbital frame following the LIGO convention.  
The following parameter ranges are covered: $\eta \in [0.05, 0.25]$, $\chi_{1,x} \in [-0.85, 0.89]$, $\chi_{1,y} \in [-0.85, 0.83]$, $\chi_{1,z} \in [-0.85, 0.99]$, $\chi_{2,x} \in [-0.85, 0.89]$, $\chi_{2,y} \in [-0.82, 0.83]$, $\chi_{2,z} \in [-0.85, 0.85]$.
Estimations of the NR data errors show that the NR final spin error is one order of magnitude smaller than the DNN errors described on Section~\ref{results}, while the remnant mass errors have comparable range~\cite{Varma:2018aht}. 
When comparing with the LALSuite algorithm library fits, the approximated precession is defined similarly to~\cite{hofmann2016final} where the input parameters are $\eta$, the total spins $\chi_{1}$ and $\chi_{2}$, the tilt angles $\theta_{1}$ and $\theta_{2}$ and the planar spin projection angle difference $\phi_{12}$ as defined in equations~(\ref{eq:tilt}) and~(\ref{eq:phi}):
\begin{equation}
  \cos \theta_{1,2} = \frac{\overrightarrow{\chi_{1,2}} \cdot \overrightarrow {L_{1,2}} }{|| \overrightarrow{\chi_{1,2}}|| \  ||\overrightarrow {L_{1,2}}|| },
  \label{eq:tilt}
\end{equation}
\begin{equation}
  \cos \phi_{12} = \frac{\overrightarrow{\chi_1^{plane}} \cdot \overrightarrow{\chi_2^{plane}}}{||\overrightarrow{ \chi_1^{plane}} || \  ||\overrightarrow{\chi_2^{plane}}||}, 
  \label{eq:phi}
\end{equation}
where $ \overrightarrow{\chi^{plane}} = \overrightarrow{\chi} - \left(\overrightarrow{\chi} \cdot \overrightarrow{L} \right) \overrightarrow{L} $ and $\overrightarrow{L} $ is the normalized angular momentum.

\begin{table}[h]
      \captionsetup{width=.85\textwidth}
      \captionsetup{font=small}
\begin{center}
  \begin{tabular}{ | l | c |  c |}
    \hline
      \textbf{NR code} &  \textbf{non-precessing} &  \textbf{precessing} \\ \hline
	 SpEC           & 592   & 1420  \\  \hline
     LazEv           & 280    & 0  \\  \hline
     MayaKranc  & 125    & 0  \\  \hline
     BAM             &  47     &  0 \\  \hline
     $\eta \rightarrow 0$             &  300     &  0 \\ \hline
     \textbf{Total}            &  \textbf{1344} & \textbf{1420}  \\  \hline
  \end{tabular}
\end{center}
\caption[Table caption text]{Summary of the NR simulations used to train the DNN in predicting the BH remnant properties. The $\eta \rightarrow 0$ row corresponds to the analytically known extreme mass ratio case described in Section~\ref{emr}. }
\label{tab:nrdata}
\end{table}

\subsection{Extreme mass ratio limit}
\label{emr}

In the extreme mass ratio limit of $\eta \rightarrow 0$, the BBH system can be approximated by a particle orbiting around a Kerr BH. 
The radiated energy and orbital momentum of the non-precessing system is known analytically~\cite{bardeen1972rotating} at the inner stable closest orbit (ISCO) as given in equations~\ref{eq:eisco} and ~\ref{eq:lisco}:
\begin{equation}
  E_{ISCO}(\eta, \chi) = \eta \left( 1 - \sqrt{\frac{2}{3 \rho_{ISCO}(\chi_f) }} \right) 
  \label{eq:eisco}
\end{equation}
\begin{equation}
  L^{orb}_{ISCO}(\eta, \chi) =\frac{2 \eta \left( 3 \sqrt{\rho_{ISCO}(\chi)} -2 \chi) \right)} {\sqrt(3 \rho_{ISCO}(\chi)  )}
  \label{eq:lisco}
\end{equation}
where $\rho_{ISCO}$ is the radius at the ISCO :
\begin{align*}
  &\rho_{ISCO}(\chi) = 3 + Z_2 - sign(\chi) \sqrt{(3-Z_1)(3 + Z_1 + 2 Z_2)} \\
  &Z_1 = 1 + (1 - \chi^2)^{1/3} \left[ (1 + \chi^2)^{1/3} + (1 - \chi^2)^{1/3} \right] \\
  &Z_2 = \sqrt{ 3 \chi^2 + Z_1^2}
  \label{eq:rhoisco}
\end{align*}
The particle plunging into the BH after the ISCO, the final radiated energy is $E_{rad} = E_{ISCO}$ leading to a final mass value of $M_f = 1- E_{ISCO}$. 
The final spin is obtained by solving numerically equation~\ref{eq:chifin}: 
\begin{equation}
  \chi_f = \frac{L_{orb} + S_1 +S_2}{M_f^2}
  \label{eq:chifin}
\end{equation}
Following the approach in~\cite{Jimenez-Forteza:2016oae}, we use the analytical results described above to add inputs in the $\eta \rightarrow 0$ limit to the NR samples shown in Table~\ref{tab:nrdata}.
We have tested adding samples of $n = {100, 300, 1000}$ points and have concluded from the analysis of the DNN errors that the value of 300 points enabled to increase the DNN accuracy in estimating the remnant parameters at large mass ratio without decreasing the accuracy for similar masses binaries.
This additional sample enables to enhance the volume of the parameter space on which the DNN is trained, as well as better extrapolate the prediction at high mass ratio where little NR simulations are available due to the high computational cost in this regime.

\section{Predicting the remnant mass and spin}
\label{fits}

\subsection{Deep Neural Network}
\label{dnn}

The DNN is built with the \texttt{keras} package of the \texttt{TensorFlow} software version 1.13.1~\cite{tensorflow2015-whitepaper}.
The non-precessing and precessing datasets described in Section~\ref{inputs} are separated into three subsets: the training, validation, and testing samples containing respectively 80\%, 10\% and 10\% of the full datasets.
Each sample spans a similar range in the parameter space and is standardized in order to obtain a mean value of 0 and a standard deviation of 1 on the training dataset to ensure a proper convergence of the algorithm.
The DNN is trained on the training samples with its hyperparameters tuned heuristically, including the number of layers, the number of nodes, the activation functions, the optimizer method, the optimizer learning rate and decay as summarized on Table~\ref{tab:dnn}. 
The best hyperparameters values were selected by comparing the prediction performance on the validation dataset, and the \texttt{keras-tuner} package performed an optimised comparison of DNN configurations with the random search option  to ensure that no more efficient configuration exists.

\begin{table}[h!]
      \captionsetup{width=.85\textwidth}
      \captionsetup{font=small}
      \begin{center}
  \begin{tabular}{ | l |  c |  c | }
    \hline
      \textbf{Hyperparameter} &  \textbf{Tested values} &  \textbf{Best value} \\ \hline
	 Number of layers             & 3,4,5,6   & 4  \\ \hline
     Nodes per layer               & 8192, 4096, 2048, 1024, 512 & 4096, 256, \\ 
                                            &  256, 128, 64, 32, 16, 8, 4    &   64, 8  \\ \hline
     Activation functions        & ReLU,   & ReLU  \\ 
                                           & LeakyReLU($\alpha=0.1,0.3,0.5$)    &   \\ \hline
     Optimizer method             &  Adam, SGD     &  Adam \\ \hline
     Optimizer learning rate     &  $10^{-3}$, $10^{-4}$, $10^{-5}$     &  $10^-3$ \\ \hline
     Optimizer decay                &  0, $10^{-3}$, $10^{-4}$ & $10^{-4}$  \\ \hline
     Adam $\epsilon$              &  0, $10^{-9}$, $10^{-8}$, $10^{-1}$     &  0 \\ \hline
     SGD momentum              &  0.1, 0.3, 0.5, 0.7, 0.9     &   not used \\ 
    \hline    
  \end{tabular}
\end{center}
\caption{The different values tested for the heuristic optimization of the DNN architecture. Not all combinations of nodes per layers were tested, but more than 50 iterations of decreasing values were probed with \texttt{keras-tuner}.}
\label{tab:dnn}
\end{table}

All DNN contains an input layer followed by four hidden layers with respectively 4096, 256, 64, 8 nodes, activated by a rectified linear unit (ReLU) function. 
The last output layer ends with a linear activation function on two nodes resulting in the prediction of the output values $M_f$ and $\chi_f$.
The loss function is the mean absolute error, minimized using the adaptative stochastic gradient optimizer Adam~\cite{kingma2014adam} with learning rate of $10^{-3}$ and decay of $10^{-4}$.
The validation sample is not only used to select the best hyperparameters but also to avoid overfitting during the training phase, that is characterized by a decreasing loss on the training sample while all the information contained in the data have been processed into tuning the DNN.
This is avoided by implementing a stopping procedure quitting the training phase when the loss is constant on the validation sample, on which the DNN is not trained.
In order to ensure a correct estimation of the DNN prediction performance, the final results shown on Section~\ref{results} are then obtained on the testing dataset that is not used during the training and validation procedures. 
As DNNs are known to have limited extrapolation outside the parameter space where they have been trained, their robustness is tested by generating $10^4$ BBHs with random initial parameters and verifying that the predicted remnant BH properties are below the Kerr limit, without finding any violation.

\subsection{Results}
\label{results}

\paragraph{Non-precessing case}

\begin{figure}[ht!]
      \captionsetup{width=.85\textwidth}
      \captionsetup{font=small}
          \centering
    \begin{subfigure}[t]{0.45\textwidth}
        \centering
        \includegraphics[width=\textwidth]{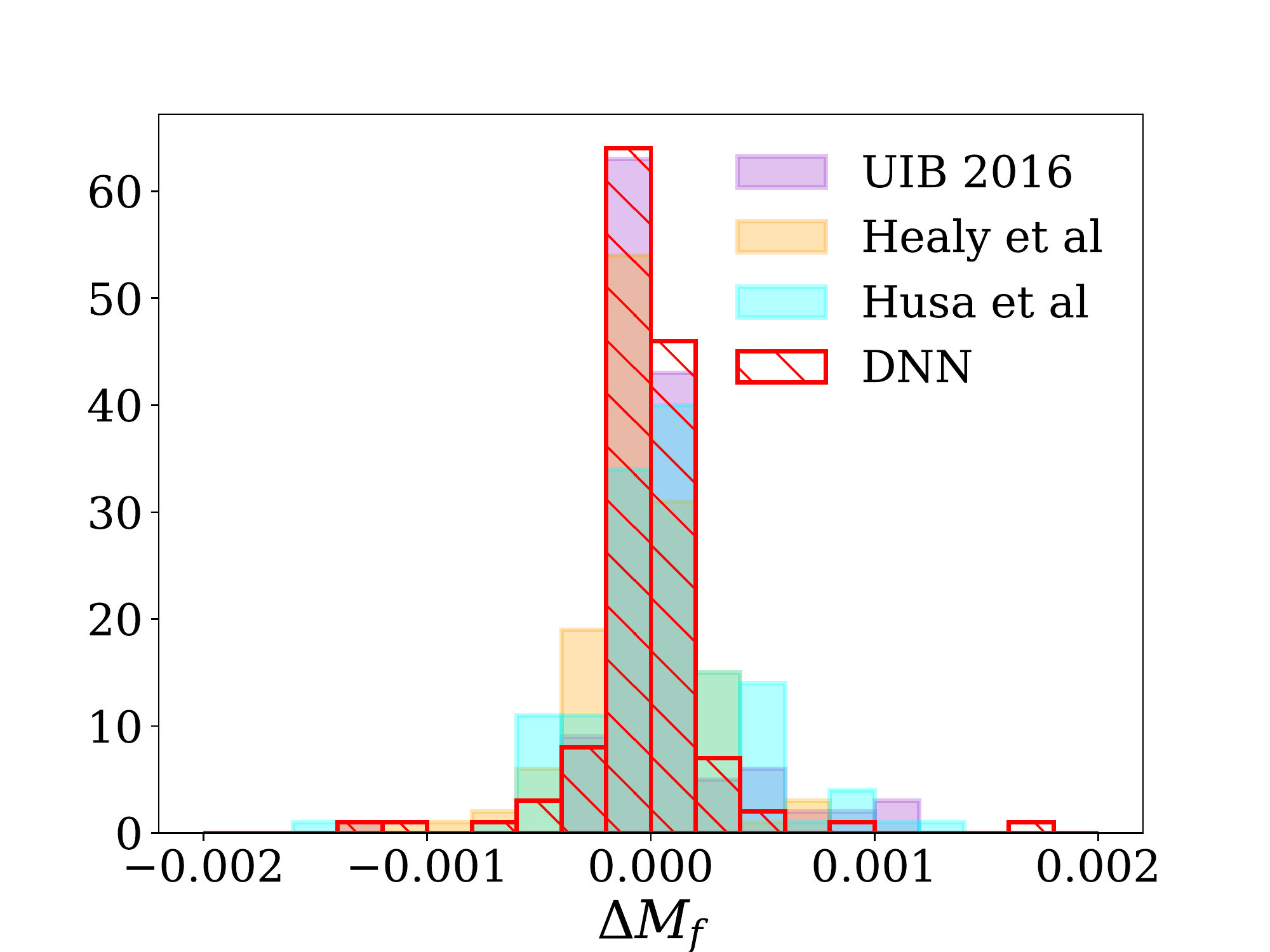}
    \end{subfigure}%
    ~ 
    \begin{subfigure}[t]{0.45\textwidth}
        \centering
        \includegraphics[width=\textwidth]{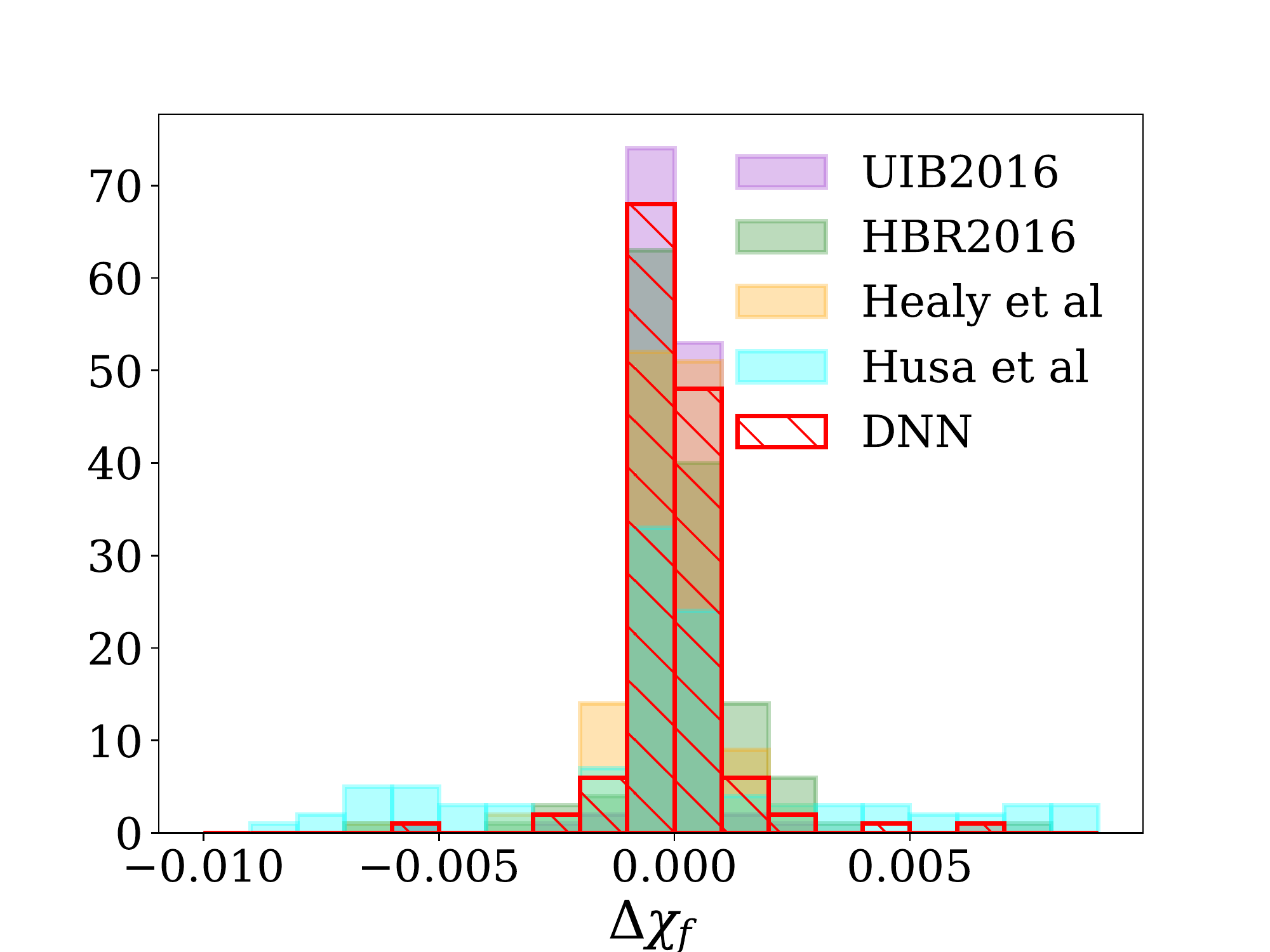}
    \end{subfigure}
    \caption{Residual error on the remnant mass $M_f(\eta, S_{eff}, \Delta \chi)$ (left) and spin magnitude $\chi_f(\eta, S_{eff}, \Delta \chi)$ (right)  as predicted by the DNN on the testing sample of non-precessing BBHs. Our error is compared with the fits performed by the UIB group in 2016~\cite{Jimenez-Forteza:2016oae}, Healy et al~\cite{PhysRevD.90.104004},  Husa et al~\cite{Husa:2015iqa} and Hofmann, Barausse and Rezzolla (HBR)~\cite{hofmann2016final}.}
   \label{fig:prederror}
\end{figure}

In a first time, we ensure that DNN correctly estimate the remnant mass and spin of non-precessing BBHs that are fully described in the existing fits by predicting the final mass and spin magnitude using the test sample not used during the training. 
We observe that the DNN correctly captures the relationship between the initial BBH parameters and the remnant properties as shown on the remnant mass and spin residual error distribution $\Delta M, \chi_f = M, \chi_f^{DNN} - M, \chi_f^{NR} $ of Figure~\ref{fig:prederror}.

\begin{figure}[ht!]
  \captionsetup{width=.85\textwidth}
        \captionsetup{font=small}
    \centering
	\includegraphics[width=0.85\textwidth]{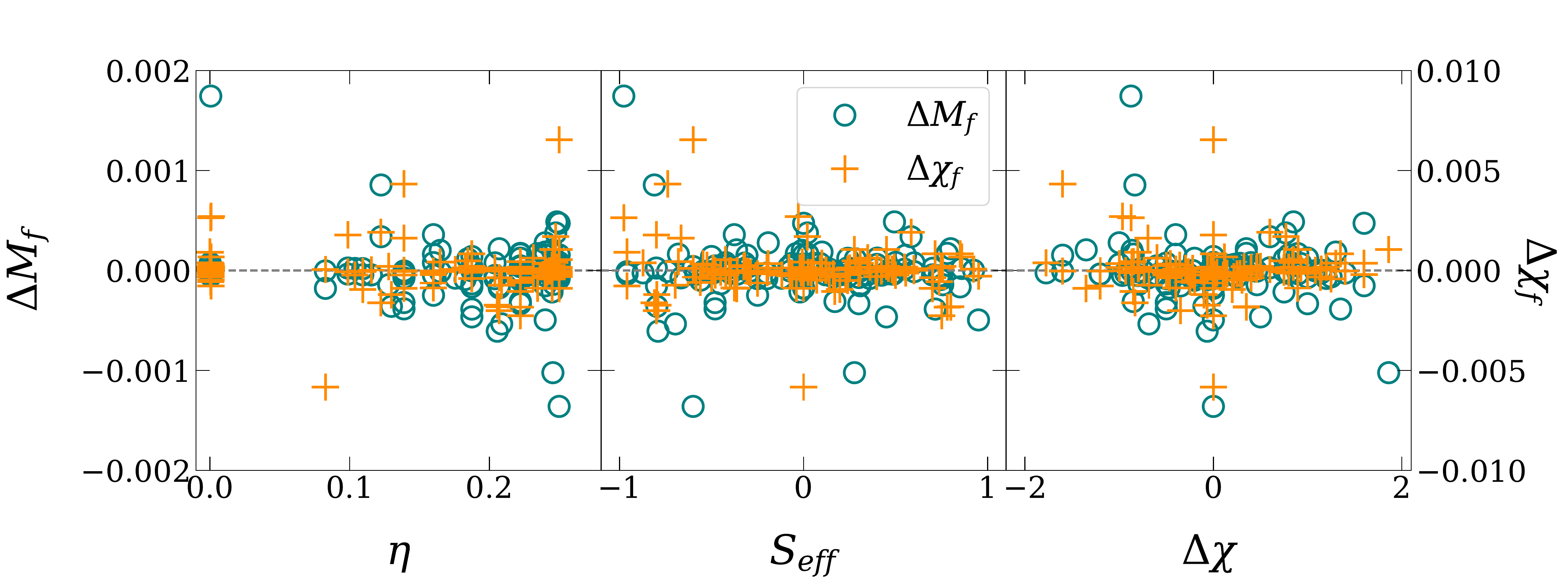}
      \caption{Distribution of the residual errors $\Delta M_f$ and $\Delta \chi_f$ as a function of the DNN inputs: the BBH mass ratio $\eta$, effective spin $S_{eff}$ and spin difference $\Delta \chi$. }
   \label{fig:prederrorin}
\end{figure}

We compare our results with the existing remnant mass and spin fits available in the \texttt{nrutils.py} code of the \texttt{LALInference} package of the \texttt{LALSuite} software used by the LIGO-Virgo collaborations~\cite{lalsuite}.
For the non-precessing BBHs, \texttt{nrutils} includes the three-dimensional remnant fits of in~\cite{PhysRevD.90.104004},~\cite{hofmann2016final} and~\cite{Jimenez-Forteza:2016oae}, as well as the two-dimensional fit in~\cite{Husa:2015iqa} where the spin difference $\Delta \chi$ is not included.
The DNN improves the prediction of the final mass $M_f$, as the residuals root mean square (RMS) of the $\Delta M_f$ distribution is $2.8 \cdot 10^{-4}$, to compare with the range $[4,5]\cdot 10^{-4}$ for the other methods shown. 
The prediction of the final spin is similar for the remnants fits and the DNN, as the residuals root mean square (RMS) is $\mathcal O( 10^{-3})$ for all three-dimensional methods, while it is found to be  $\mathcal O( 10^{-2})$  for the two-dimensional fit in~\cite{Husa:2015iqa}, indicating that the initial spin difference impacts the final spin value. 
The distribution of the DNN errors as a function of the input parameters shown on Figure~\ref{fig:prederrorin} does not show a specific trend towards a specific region of phase-space, and while the largest errors correspond to points with extreme $\eta$ values, we also note that this area of the parameter space is the most populated with the majority of events displaying low errors.

\begin{figure}[ht!]
      \captionsetup{width=.85\textwidth}
      \captionsetup{font=small}
          \centering
    \begin{subfigure}[t]{0.45\textwidth}
        \centering
        \includegraphics[width=\textwidth]{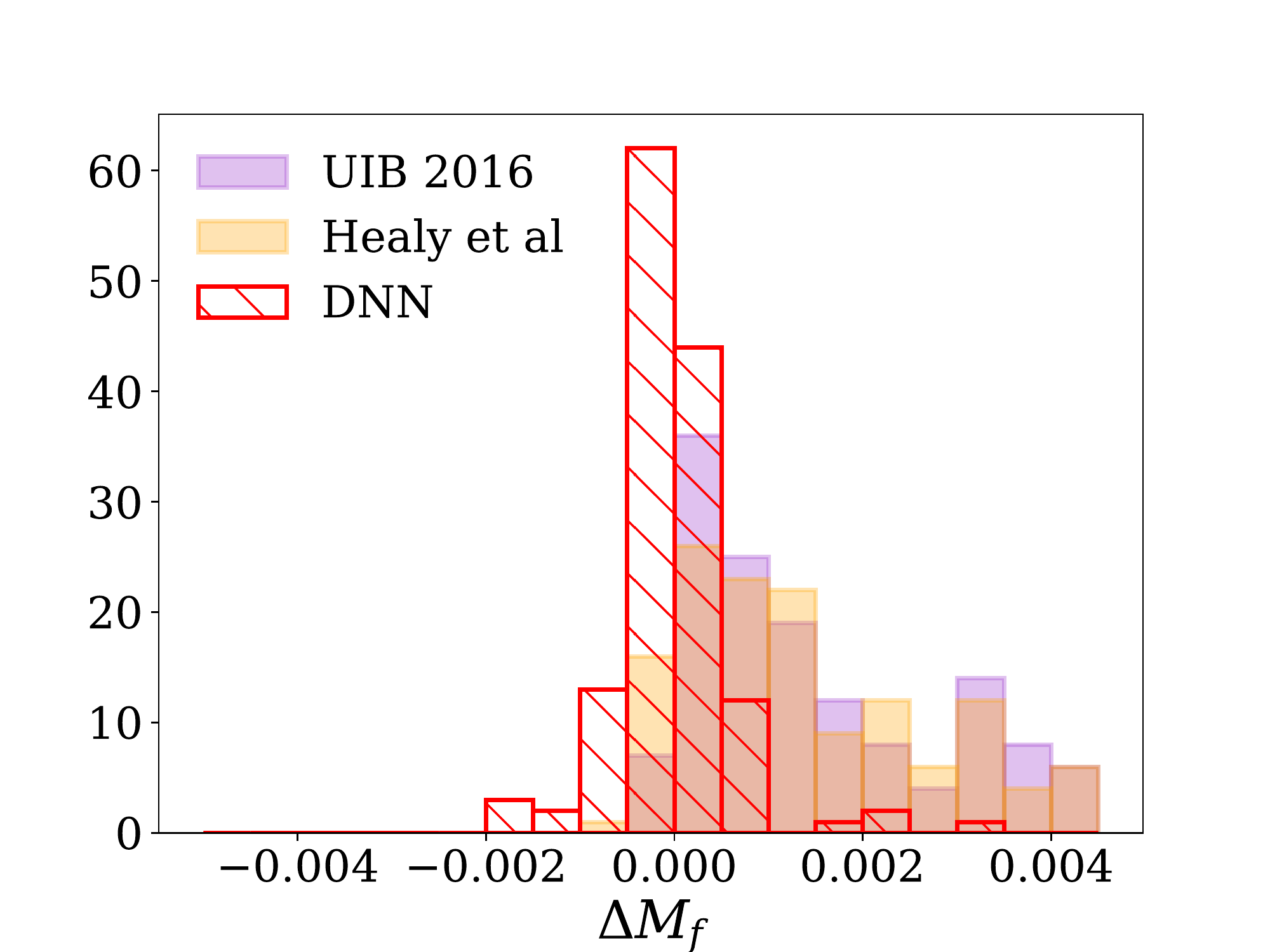}
    \end{subfigure}%
    ~ 
    \begin{subfigure}[t]{0.45\textwidth}
        \centering
        \includegraphics[width=\textwidth]{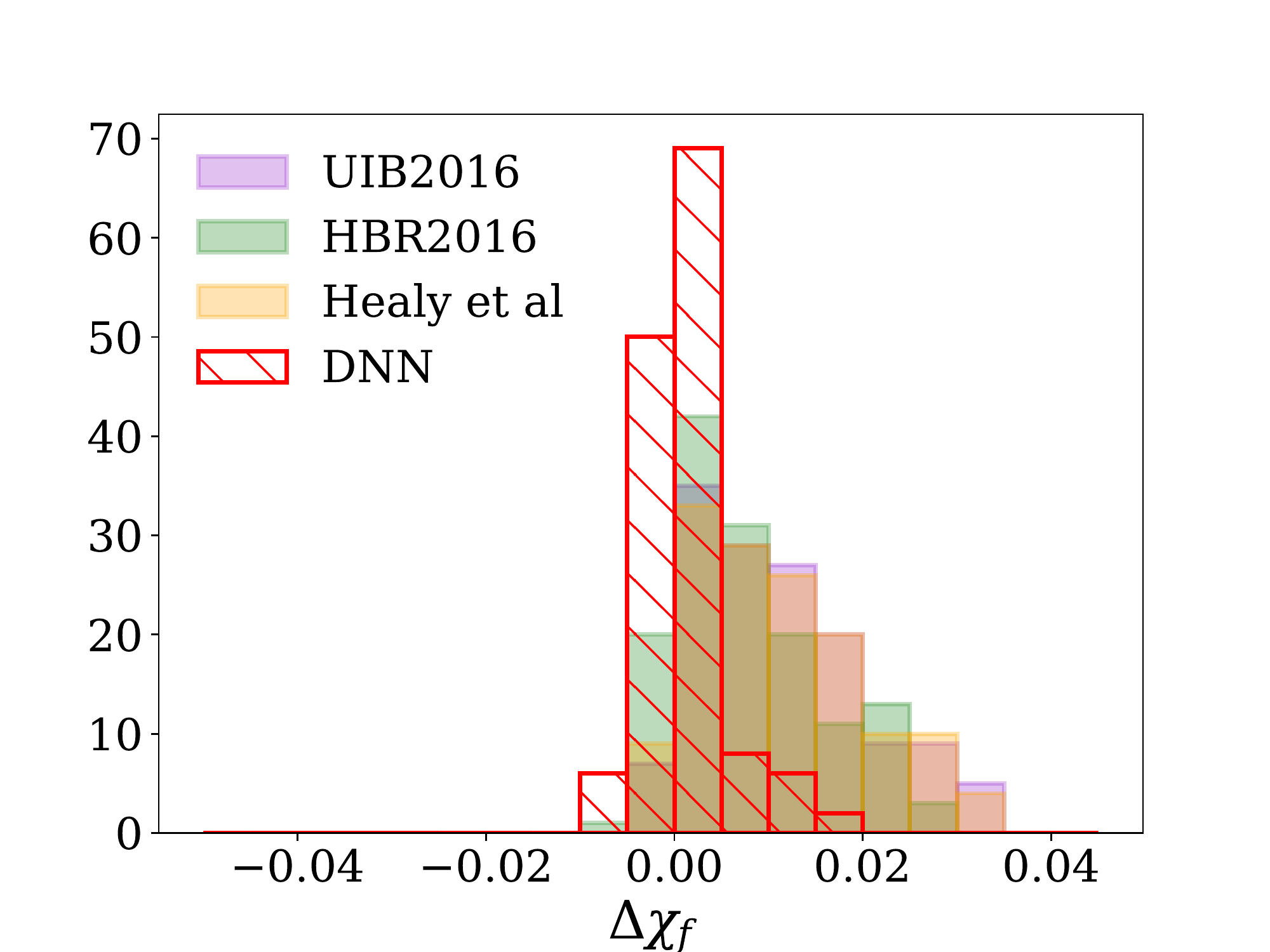}
    \end{subfigure}
    \caption{Residual error on the remnant mass $M_f(\eta, \chi_{1x,y,z}, \chi_{2x,y,z})$ (left) and spin magnitude  $\chi_f(\eta, \chi_{1x,y,z}, \chi_{2x,y,z})$ (right)  as predicted by the DNN for the testing sample of precessing BBHs. Our error is compared with the fits performed by the UIB group in 2016~\cite{Jimenez-Forteza:2016oae}, Healy et al~\cite{PhysRevD.90.104004} and Hofmann, Barausse and Rezzolla (HBR)~\cite{hofmann2016final}.}
   \label{fig:prederrorprecchif}
\end{figure}

\paragraph{Precessing case}
In a second time, we test the generalization of the DNN to the 7-dimensions parameter space of precessing binaries.
This analysis presents a generalization of the remnant mass fits to precessing binaries that is not available in the \texttt{nrutils} package, where only aligned-spins fits are available for the final mass.
The package includes remnant spin fits for precessing BBHs based on  aligned-spins binaries "augmented" with the in-plane spin contribution added to the final spin, as in~\cite{Jimenez-Forteza:2016oae},~\cite{PhysRevD.90.104004} and~\cite{Husa:2015iqa}, and in one case an additional parameter captures the precession dynamics~\cite{hofmann2016final}.
We compare their prediction with the accuracy of the DNN using the testing sample of our precessing catalog on the $\Delta M, \chi_f = M, \chi_f^{DNN} - M, \chi_f^{NR} $ distributions of Figure~\ref{fig:prederrorprecchif}.
We observe that including the full spin vector as an input to the DNN enables to determine the remnant parameters with better accuracy than fits based on approximating the precession, that tends to overestimate the final quantities.
The improvement can be observed in the RMS of the error distribution being decreased by a factor two for the DNN compared to the existing fits, it is $8 \cdot 10^{-4}$ for $M_f$ and $4 \cdot 10^{-3}$ for $\chi_f$.
In total, more than 90\% of the $M_f$ and $\chi_f$ predictions on testing samples are below 0.1\% and 1\% respectively, against 67\% for~\cite{hofmann2016final} and less than 50\% for~\cite{Jimenez-Forteza:2016oae} and~\cite{PhysRevD.90.104004}.

\begin{figure}[ht!]
 \captionsetup{width=.85\textwidth}
      \captionsetup{font=small}
   \centering
	\includegraphics[width=0.85\textwidth]{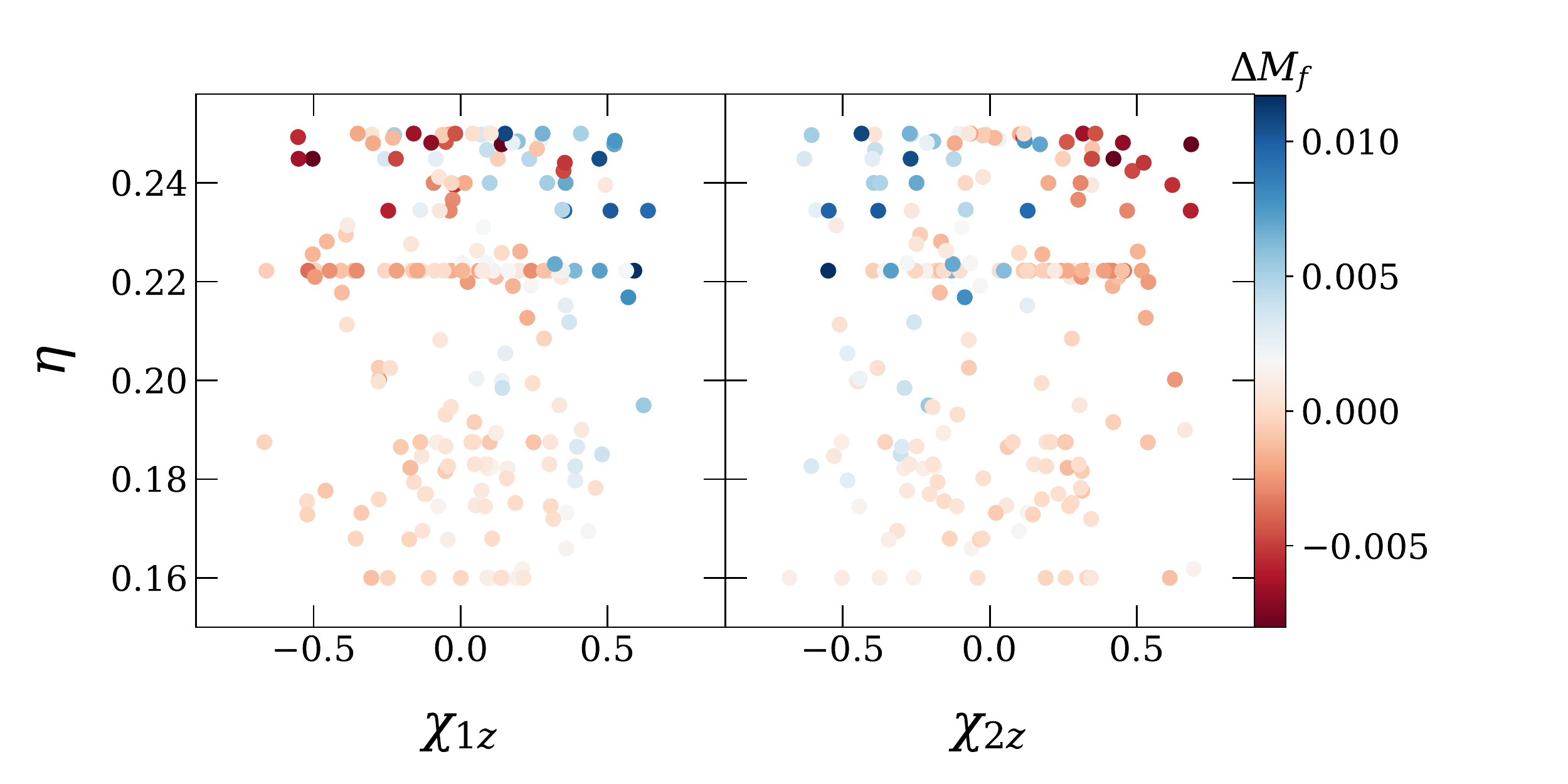}
      \caption{Distribution of the residual errors $\Delta M_f$ as a function of the DNN inputs: the BBH mass ratio $\eta$, and the spin components $ \chi_{1z}$ and $ \chi_{2z}$ when using the sample on randomly generated BBH simulations. }
   \label{fig:predprecerrorin}
\end{figure}

Similarly to the non-precessing case, we find that the error obtained on the testing sample are homogenously distributed across the parameter space.
In order to characterize the ability of the DNN to generalize outside the sample it was trained on and study its accuracy on a large parameter space, we generate 200 random BBH configuration for which we determine the final mass with the \texttt{surfinBH} package providing interpolation between NR waveforms~\cite{Varma:2018aht}.
We observe on Figure~\ref{fig:predprecerrorin} that the error on the final mass generally remains at the subpercent level and is larger at high $\chi_{1,2z}$ values, indicating that the DNN is less efficient in capturing the large precession at the edge of the parameter space. 

\section{Conclusion}
\label{conclusion}

We demonstrate that DNNs trained on NR data are a efficient method to predict the mass and spin of the remnant BH in a BBH merger, which suggests that DNN methods can be used in a similar way to more general applications in waveform modelling, e.g. to model the coefficients of a phenomenological waveform model across the parameter space, including in the presence of precession.
The performance of our DNN is validated by comparing its accuracy with the existing fits available in the public library of the LIGO software, where it achieves a twice more accurate prediction of the remnant mass and a similarly accurate prediction of the final spin for non-precessing binaries.
The DNN is shown to be specifically powerful in generalizing the remnant prediction to pecessing binaries, providing an estimate on the final mass and spin with errors smaller $10^{-3}$ and $10^{-2}$ respectively with a considerably improved accuracy compared to the available explicit fits.
While the current LIGO and Virgo measurements of the remnant BH masses and spins are limited by statistical errors larger than our results, future ground- and space-based interferometers will require accurate estimates to decrease the systematical error. 

The optimization of the DNN has shown that the final results have little dependency on the hyperparameters, implying that the current limitations of the prediction are due to the limited size of the NR catalogs.
While our current analysis already spans a large parameter space by using all the publicly available NR catalogs, the DNN accuracy will certainly be improved as more NR simulations become available for training.

\section{Acknowledgements}
\label{acknowledgements}

We thank Nathan Johnson-McDaniel for advice on using the LIGO software library. 
We thank Vijay Varma for advice on using the \texttt{surfinBH} package. 
We thank the SXS, LazEv, MayaKranc and BAM collaborations for releasing and sharing their catalogs of numerical simulations of GW.
LH is funded by the Swiss National Science Foundation Early Postdoc Mobility Grant 181461.
This work was supported by European Union FEDER funds, the Ministry of Science, 
Innovation and Universities and the Spanish Agencia Estatal de Investigación grants FPA2016-76821-P,        
RED2018-102661-T,    
RED2018-102573-E,    
FPA2017-90687-REDC,  
Vicepresid`encia i Conselleria d’Innovació, Recerca i Turisme, Conselleria d’Educació, i Universitats del Govern de les Illes Balears i Fons Social Europeu, 
Generalitat Valenciana (PROMETEO/2019/071),  
EU COST Actions CA18108, CA17137, CA16214, and CA16104. 
The authors are grateful for computational resources provided by the LIGO Laboratory and supported by the National Science Foundation Grants PHY-0757058 and PHY-0823459 and the STFC grant ST/I006285/1.

\bibliographystyle{unsrt}
\bibliography{article} 

\end{document}